\documentstyle[aps,epsf,epsfig,twocolumn,psfig,array]{revtex}

\begin{document}

\draft \preprint{}

{\bf Comment on "Quantum Interferometric Optical Lithography:
Exploiting Entanglement to Beat the Diffraction Limit"
}
\\

In a recent Letter~\cite{Boto00} it was suggested to exploit the
entanglement of photons to concentrate $N$-photon absorption
spotsizes below half the wavelength $\lambda$ of the employed
light, thus beating the diffraction limit. The obvious application
of this idea is known as 'quantum lithography' for which the
description of an associated $N$-photon absorption process is
modelled by the dosage operator $ \delta_N = {(\hat e^\dagger)^N
(\hat e)^N / N! } $~\cite{Boto00}. However, the rates for such
spatially concentrated $N$-photon absorption processes are
critically low, as is shown in this Comment.

In reference~\cite{Boto00} it was concluded that "the $N$-photon
absorption cross section, with $N$ entangled photons, scales
similar to $I$ and not $I^N$ expected classically"~\cite{Boto00}.
But the authors did not mention that there is a very low threshold
intensity $I_c$, which must not be exceeded.

The absorption process is a coherent $N$-photon process, one
therefore has to avoid the presence of any photons other than
those forming the specifically designed quantum states. Their
associated coherence volume is $A c\tau$, where $A$ is the wave
packet cross section and $\tau$ its coherence time. For the
two-photon case there consequently exists a critical intensity
value $I_c=\hbar \omega / A \tau$ above which uncorrelated
multi-photo absorption processes start to take
over~\cite{Javanainen90}. Analogously, for general $N$-photon
quantum states the maximum flux is $N$ photons per coherence time,
above it, competitive processes from non-entangled photons will
contribute to the multi-photon absorption rate~\cite{Perina98},
drastically degrading the quantum lithography schemes'
performances. Hence, the intensity threshold~$I_c$ implies that
{\em only one $N$-photon wave packet may arrive per coherence time
$\tau$}.

Note, that there is an extension to the quantum lithography
proposal using squeezed states~\cite{Agarwal}. Having components
which are unbounded in photon number, these states show no
critical intensity behaviour, as those of
proposals~\cite{Boto00,{Bjoerk.PRA01}} but the image contrast also
degrades with increasing intensity.

We now want to find out what the chances are that all those $N$
photons in one coherence volume hit the same small absorber.

In~\cite{Boto00} it is said: "Recall that the photons are
correlated in space and time, as well as number. Hence, if the
optical system is aligned properly, the probability of the first
photon arriving in a small absorptive volume of space-time is
proportional to $I$. However, the remaining $N - 1$ photons are
constrained to arrive at the same place at the same time, and so
each of their arrival probabilities is a constant, independent of
$I$."

Fortunately, there is no such constraint, if there was, we could
not concentrate the photons below the classical diffraction limit
since the arrival probability of the first photon follows the
classical intensity distribution $I \propto \langle \delta_1
\rangle$ and we would, because of the assumed arrival constraint
$\langle \delta_1 \rangle \propto \langle \delta_N \rangle$, find
$\langle \delta_N \rangle \propto I$ which contradicts all the
other beautiful results reported in reference~\cite{Boto00}. For
an explicit counterexample use the number entangled state $|\psi_N
\rangle = (|N\rangle_C|0\rangle_D + e^{i N
\varphi}|0\rangle_C|N\rangle_D )/\sqrt{2}$ of
reference~\cite{Boto00}: for $K < N$ we find $ \langle \delta_K
\rangle$ is independent of $\varphi$, whereas $ \langle \delta_N
\rangle \propto 1 + \cos (2 N \varphi)$ shows the desired
multi-photon periodicity factor $N$ in the argument of the
interference pattern. We are led to conclude that the photons are
not spatio-temporally constrained as claimed in the above
citation.

That the photons are in general not spatio-temporally constrained
in the way suggested by the above citation is also indicated by
the absorption curves of $\langle \delta_1 \rangle \ldots \langle
\delta_4 \rangle$ for four photon reciprocal binomial states
displayed in FIG.~7 of reference~\cite{Bjoerk.PRA01}.

In order to estimate an upper limit on the photons' joint arrival
probability, let us assume (unrealistically) that the photons
could somehow be concentrated below the classical diffraction
limit and all be made to impinge upon an area corresponding to the
minimal spotsize $(\lambda/2N)^2$ resolvable by quantum
lithography. Even this area is still orders of magnitude larger
than a single molecule's absorption cross-section. The ratio of
spotsize (for 200 nm light) of $N^{-2} \cdot 10^{-14} \mbox{m}^2$
compared to a large single-photon absorption cross-section of less
than $ 10^{-19} \mbox{m}^2$ leads to a relative coverage by the
molecule of $r\approx N^2 \cdot 10^{-5}$ at most. Having absorbed
the first photon the molecule's chance of absorbing the following
$N-1$ photons is less than $r^{N-1}$. This discussion has not
included the effects due absorption suppression in non-resonant
multi-photon absorptions. In either case the low absorption
probability seems to render the quantum lithography proposal in
its present form unviable.
\\ \\
{O. Steuernagel}, {Dept. of Physical Sciences, University of
Hertfordshire, College Lane, Hatfield, AL10 9AB, UK}
\\
ole@star.herts.ac.uk
\\

\end{document}